# Data Mining for Terahertz Generation Crystals

Gabriel A. Valdivia-Berroeta, Zachary B. Zaccardi, Sydney K. F. Pettit, (Enoch) Sin-Hang Ho, Bruce Wayne Palmer, Matthew J. Lutz, Claire Rader, Brittan P. Hunter, Natalie K. Green, Connor Barlow, Coriantumr Z. Wayment, Daisy J. Harmon, Paige Petersen, Stacey J. Smith, David J. Michaelis*, Jeremy A. Johnson*

Department of Chemistry and Biochemistry, Brigham Young University, Provo, UT 84602, USA

* Electronic Mail: dmichaelis@chem.byu.edu, jjohnson@chem.byu.edu

We demonstrate a data mining approach to discover and develop new organic nonlinear optical crystals that produce intense pulses of terahertz radiation. We mine the Cambridge Structural Database for non-centrosymmetric materials and use this structural data in tandem with density functional theory calculations to predict new materials that efficiently generate terahertz radiation. This enables us to (in a relatively short time) discover, synthesize, and grow large, high-quality crystals of four promising materials and characterize them for intense terahertz generation. In a direct comparison to the current state-of-the-art organic terahertz generation crystals, these new materials excel. The discovery and characterization of these novel terahertz generators validates the approach of combining data mining with density functional theory calculations to predict properties of high-performance organic materials, potentially for a host of exciting applications.

1. Introduction

   The discovery and design of advanced solid materials with useful properties and applications is essential to the advancement of many fields, including spectroscopy, catalysis, electron transport, energy storage and release, and air and water purification. As a subset in solid materials research, organic materials provide a powerful advantage in that desired material properties can be finely tuned by editing the structure of molecular building blocks. For organic crystals, the function of the material is governed by the chemical and photophysical properties of the molecular building blocks, along with molecular packing, molecular orientation in the crystal, and specific surface geometries. Therefore, the optimization of organic material properties is complicated by the need to control both molecular properties and solid-state packing preferences. Materials databases, like the Cambridge Structural Database (CSD), compile and make available a wealth of structural information for materials that have been developed for specific applications [1]. One potential use of such databases is to identify already existing materials that may be ideal for applications other than their original intended use. The idea that one can *easily* mine information about known materials for the development of new and extremely useful purposes will rapidly accelerate the discovery of new materials for many applications. This data mining approach also gives rise to new screening methodologies in the rapidly growing field of materials informatics [2-7].

Various data mining approaches have been reported that classify material properties for a host of applications [8-12]. In addition, while the data mining and classification of inorganic materials is increasingly common, approaches to identify organic materials are rare [13-15]. However, most of these reports for organic and inorganic materials give only theoretical predictions and lack experimental validation. For example, in Ref. [14], they used data mining in combination with first-principles calculations of density of states to identify new candidate high-$T_c$ superconductors, however, no candidate materials were actually tested for superconductivity. In this report we combine data mining of known organic materials from the CSD with computational analysis of key molecular properties to identify new candidate organic materials for intense terahertz (THz) generation. We then validate our combined data mining and computational approach to materials discovery by synthesizing and fully characterizing four new THz generating organic materials.

Terahertz radiation, with frequencies from 1-10 THz (wavelengths of 300-30 μm), exhibits unique interactions with many materials and thus is able to analyze and control material properties in ways that differ from other forms of radiation. Many emerging and potentially disruptive applications of THz spectroscopy are taking advantage of these unique interactions, including in bioimaging and security [16], chemical recognition [17], non-destructive chemical monitoring in industry and food processing [18,19], and wireless communication and high-speed computational devices [20]. Of the various methods available for generating THz light, optical rectification of infrared (IR) light with organic nonlinear optical (NLO) crystals is the most efficient method to produce high-intensity THz fields with extremely broad bandwidths (**Figure 1a**) [21-28]. NLO crystals made up of organic molecular building blocks hold significant promise for THz generation due to the ease of editing and creating new organic chromophores with custom properties. However, very few efficient organic NLO crystals for THz generation have been reported and even fewer crystals are commercially available [24,27,29,30]. This currently limits the potential applications of high-power THz spectroscopy. The discovery of novel, high-performing organic NLO materials will impact the efficiency of intense THz generation and will open and expand avenues for NLO applications that utilize harmonic generation, optical parametric amplification, optical phase conjugation, and electro- and acousto-optic modulation.

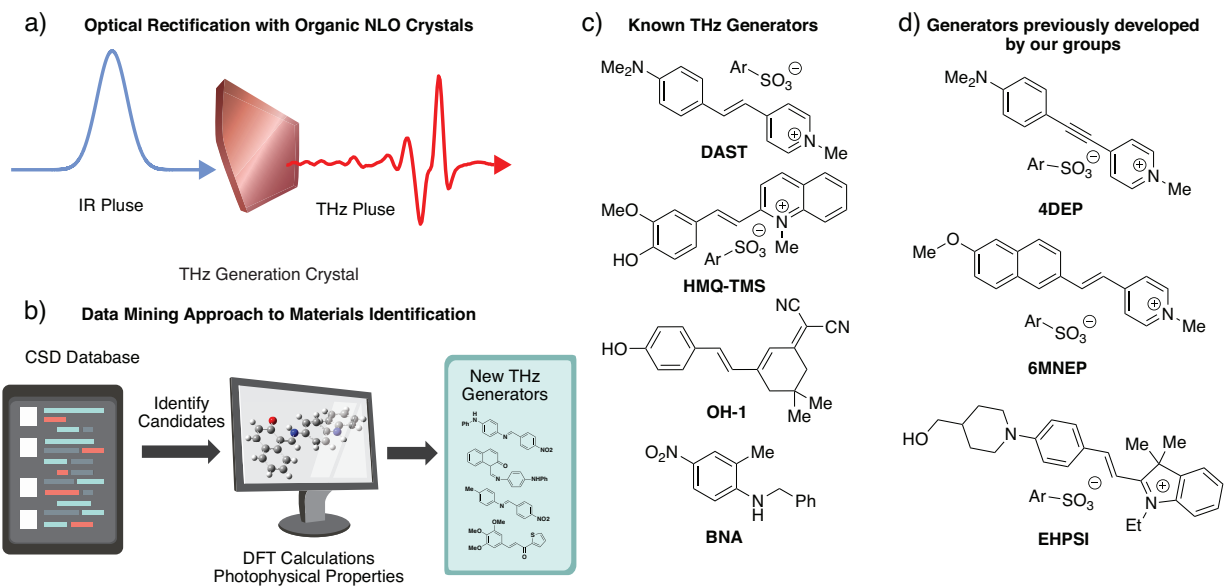

**Figure 1**. a) Optical rectification scheme for conversion of IR light to THz frequency radiation. b) Combined data mining/computational approach to identification of new organic NLO crystals for intense THz generation. c) Structure of known, state-of-the-art organic NLO crystals. d) Organic NLO crystals previously developed by our groups via modification of known generator structures.

The classical approach to develop new organic materials for intense THz generation is a tedious process that involves both molecular design, crystal structure determination, and crystal growth optimization. The first step involves identification of new organic chromophores with ideal photophysical properties. Promising molecules must then form large single crystals with a suitable molecular packing that aligns chromophores in a parallel fashion. In this work, we greatly accelerate this development process by using structural data mining tools to identify known organic materials from the CSD with ideal non-centrosymmetric crystal properties for THz generation (**Figure 1b**). We then perform first-principles calculations of the core molecular property that influences nonlinear optical properties (hyperpolarizability), and we inspect crystal packing to rank the molecular crystals based on their potential for THz generation. Through these efforts, we identify several promising organic materials with previously unknown THz generation properties and report the synthesis and crystal growth of large, high-quality crystals for four of the most promising candidates. We also characterize these four new materials for their THz generation capabilities and compare with benchmark organic THz generators. Our new organic material *(E)-4-((4-nitrobenzylidene)amino)-N-phenylaniline* (PNPA) shows exceptional THz generation properties that exceed current state-of-the-art red-orange organic THz generators such as DAST and OH-1 (**Figure 1c**) [24]. We also discover a yellow THz generator (E)-4-((4-methylbenzylidene)amino)-N-phenylaniline (NMBA) that compares favorably with yellow THz generator BNA, which has recently gained attention as a new standard for THz generation [29,31-33].

**Results and discussion:**

Previous approaches to developing organic NLO crystals from us [25-28,34] and others [21,35-44] have shown that the process to identify, crystalize, and develop new organic THz generators can take years. Even when these efforts are successful in accessing new NLO crystals, the improvements in THz output are often small compared to benchmark organic crystals like DAST, HMQ-TMS, OH-1, and BNA (**Figure 1c**). Our previous efforts in organic NLO design have resulted in the development of new THz generation crystals, including 4DEP [26], P-BI [25], EHPSI-4NBS [27,30], and 6MNEP [28] (**Figure 1d**). The new chromophores 4DEP, EHPSI, and 6MNEP resulted from modifications to the known structure of NLO crystals P-BI and DAST. EHPSI, the highest performing of the three, outperforms state-of-the-art NLO crystal OH-1 by generating stronger peak THz electric fields with a broader spectrum [30]. However, EHPSI required years of development and only resulted in a ~1.2 times improvement in THz output. In addition, our previous work, as with most recent work in THz crystal development, relied on modification of previously known chromophores for THz generation and did not lead to the discovery of new classes of THz generating chromophores [25-28,30]. In contrast, our data mining efforts reported here have led to the rapid discovery and development of four new THz generators with unique chromophores in a short amount of time (<1 year) that rival the efficiency of previously reported generators.

To optimally generate THz light, organic crystals must possess two key parameters: the molecular units of the crystal must have high hyperpolarizability and the molecules must align in the crystalline state in a non-centrosymmetric head-to-tail arrangement. Hyperpolarizability is a molecular property that relates to how the dipole moment of a molecule responds or is polarized by the electric field of incoming pump light. Molecules that contain highly conjugated systems of electrons often have high hyperpolarizability because of the large size of the polarizable electron cloud. In this study we use DFT calculations to quantify the hyperpolarizability of candidate molecules as an additional screening method for identifying the most promising candidates for THz generation. The second requirement is that molecules pack in a perfectly parallel head-to-tail arrangement in the crystal state that aligns their hyperpolarizability vectors. Because the CSD contains crystal structures of organic materials, we can screen candidate organic molecules for THz generation based on large hyperpolarizabilities and their crystalline alignment. The combination of these two properties allows us to identify highly promising new organic materials for THz generation.

Our initial data mining efforts to identify new organic materials involved use of an automated search of the CSD for known organic crystals. As noted above, our search focused on compounds that contain conjugated π-systems, and thus likely possess high molecular hyperpolarizability. A custom python program was developed to search the CSD 2020.0 database that included several filters to find organic materials with optimal properties. First, due to the ease of processing neutral organic crystals (such as BNA) compared to ionic organic crystals (like DAST), we only searched for neutral molecules with C, N, S, O, P, F, Cl, or Br atoms. Only materials with *r* refinement values less than 15% were considered, indicating a reliable x-ray structure. Non-centrosymmetric crystals were selected in the next stage because of the asymmetric molecular packing required for NLO applications. Relatively small molecules (less than 600 g/mol) were considered for the following stage to keep the density number in the crystal packing as high as possible. Following this same consideration, crystal structures with only one molecular species were considered, ruling out cocrystals. We also removed chiral compounds because many non-centrosymmetric chiral structures have poor alignment of THz generating chromophores.

The chromophores from the resulting 15,782 compounds were submitted for DFT hyperpolarizability calculations and ranked accordingly (see Supporting Information for details). Previous data mining approaches to materials discovery that combined calculations of materials properties used calculations of electronic properties such as band gap [45]. To understand THz generation potential, we needed to calculate advanced (nonlinear) electronic properties, in particular, the hyperpolarizability. This ability to combine calculations of advanced properties with data mining will benefit every realm of materials development because it demonstrates that any computationally accessible molecular property can be used in materials screening. From our molecular calculations of hyperpolarizability, we selected the top ~200 candidate non-centrosymmetric crystal structures and visually inspected them to determine how nearly the chromophores approached optimal alignment. Molecular crystals with good crystal packing in tandem with molecular hyperpolarizability ($\beta_{tot}$) greater than $50 \times 10^{-30}$ esu were identified as promising for organic synthesis. **Figure 2** shows the most promising molecules identified via our data mining approach (see Supporting Information for a more complete list of calculated hyperpolarizabilities together with CSD IDs). Many of these chromophores have dramatically higher calculated hyperpolarizabilities than the commercial state-of-the-art crystal OH-1 ($\beta_{tot} = 93 \times 10^{-30}$ esu) [35], and the crystal structures exhibit good molecular alignment. In addition, many of these molecules contain unique chromophores for THz generation that have not previously been explored. While some of the materials identified were first developed for nonlinear optical applications, many were developed for completely different purposes. This latter result highlights the importance of our data mining approach in identifying completely new molecular structures and chromophores for the discovery of THz generating materials.

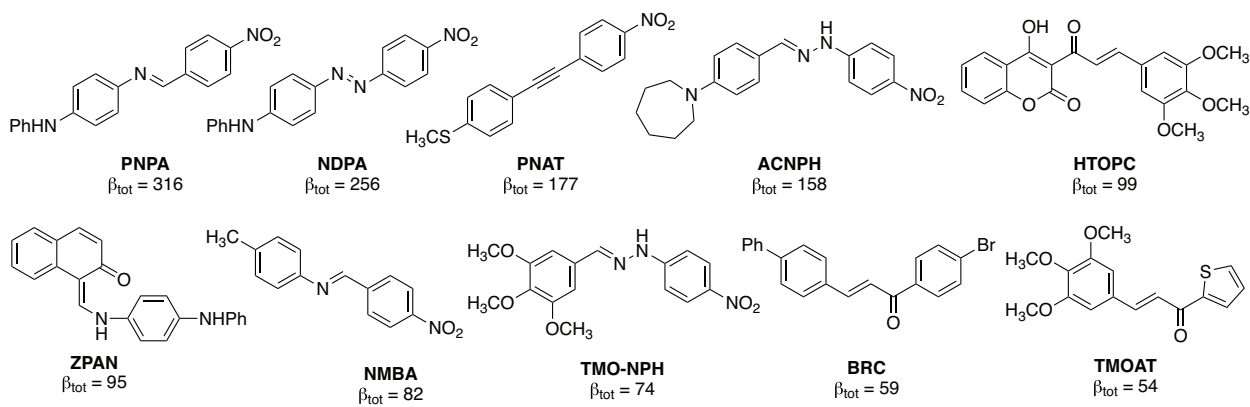

**Figure 2.** Newly identified THz generation molecules with calculated molecular hyperpolarizabilities ($\beta_{tot}$).

The synthesis of all 10 molecules represented in **Figure 2** was undertaken based on published reports. Of these molecules, four could be synthesized on a large scale and were amenable to large crystal growth, including PNPA, ZPAN, NMBA, and TMOAT (see SI for full synthesis and crystallization details). Large crystals were grown via slow evaporation protocols and representative examples of crystals grown for each molecule are shown in **Figure 3a**. X-ray crystallographic analysis was performed on each crystal to verify that the structures of the grown crystals matched the previously reported structures in the database. See Table S1 in the Supporting Information for details on X-ray analysis. The structures of all four crystals are consistent with the reported structures in the database. **Figure 3b** shows the crystal unit cell for each of the four crystals with the blue arrows indicating the direction of hyperpolarizability. It is important to note that PNPA has perfect alignment of hyperpolarizability vectors, which maximizes its potential for THz generation. We then measured the THz generation spectrum for each crystal and compared their THz generation efficiency. The unique Fourier spectrum for each crystal is shown in **Figure 3c**. Importantly, each new crystal shows THz generation at 1450 nm irradiation and can generate intense THz pulses.

Our final goal was to compare the THz efficiency and spectra of our four new THz generators to known, state-of-the-art organic NLO crystals. We separated the crystals into two different groups based on color (red crystals and yellow crystals). In the red crystal category, we compared THz generation from PNPA, ZPAN, DAST, and OH-1 by irradiation at 1450 nm pump wavelength. **Figure 4a** shows the comparison of the time traces and **Figure 4b** shows the Fourier transforms of the four red crystals when irradiated under the exact same conditions and using the same parameters for each measurement. The THz spectrum and efficiency of PNPA rivals and even exceeds both DAST and OH-1, with a spectral peak at 1 THz and frequency components extending to 5 THz. Despite the mild absorption at 2 THz, PNPA provides the benefits of large amplitudes at <1.5 THz compared to DAST, as well as more intense THz at >4 THz compared to OH-1. ZPAN, although notably less powerful, has a broad spectrum with only small absorptions before 3 THz. ZPAN does not exhibit the head-to-tail chromophore configuration, but rather the X-configuration that has been garnering recent attention due to potentially large off-diagonal components of the nonlinear susceptibility [25,26,38].

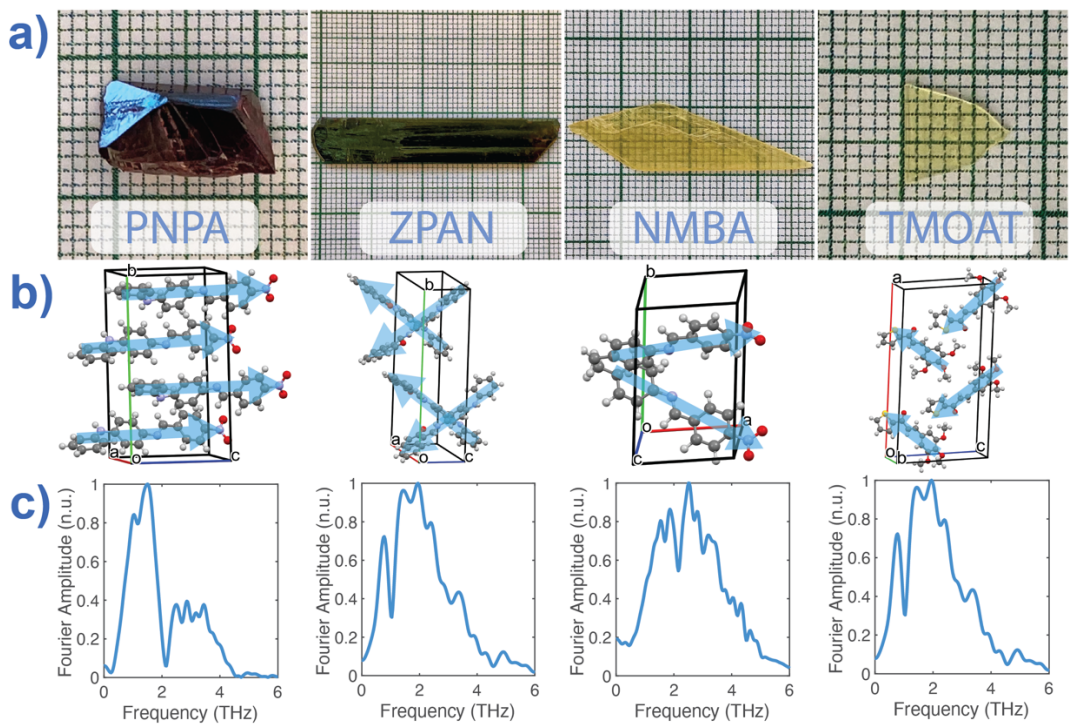

**Figure 3.** a) Large single crystals grown via slow evaporation protocols (square = 1 mm$^2$). b) X-ray crystal structure and direction of hyperpolarizability vector for each new crystal. c) Normalized THz spectrum of each NLO crystal.

Yellow organic crystals often exhibit the ability to be pumped efficiently at shorter wavelengths than their red counterparts just described. NMBA and TMOAT have nearly clear, but still slightly yellow colors. We compared THz generation using our new NMBA and TMOAT crystals to yellow BNA and the red inorganic crystal GaP. **Figure 4c** shows the electric field traces of the three yellow organic crystals compared to GaP, all pumped with exactly the same conditions, including the 1250-nm pump wavelength. **Figure 4d** shows the Fourier transforms of these four crystals, showing that THz generation using NMBA is very broadband, and rivals that of BNA. TMOAT does not generate THz as efficiently as BNA or NMBA but has a broad spectrum with only one main absorption at 1 THz. The intensity of THz generated with TMOAT is consistent with its lower calculated hyperpolarizability compared to NMBA. These results confirm that our combined data mining and computational approach can identify new materials capable of intense THz generation with high levels of success, and we can even predict relative THz generation efficiencies.

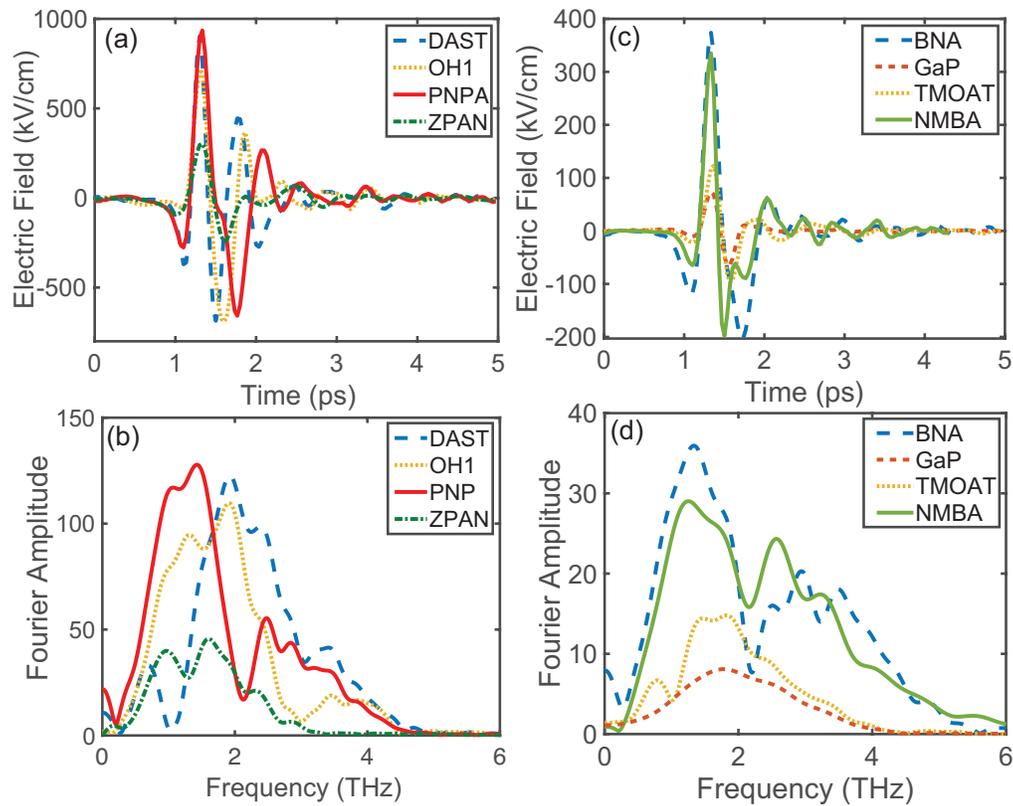

**Figure 4.** a) Generation comparison of red organic NLO crystals at a pump wavelength of 1450 nm. b) Spectra of red crystals with normalized inset. c) Generation comparison of yellow crystals pumped at 1250 nm. d) Spectra of yellow crystals with normalized inset

In conclusion, our data mining approach enables the rapid identification of new organic materials suitable for intense THz generation. These results demonstrate that data mining of structural databases combined with DFT calculations provide a productive and powerful method for identifying new organic materials for any material science application. Through this approach, unique molecular structures (chromophores) and materials can be identified that were originally reported for different materials applications, but that can be applied to a new field and enable new materials discovery. We rapidly identified and crystalized four new organic NLO crystals with unique molecular structures in the amount of time that would previously have been required to discover a signal new NLO material. In particular, the discovery of PNPA as an efficient THz generator that outperforms several benchmark organic nonlinear optical crystals will be beneficial to the growing research area of high-field THz science. These results coincide with the goals of the materials genome initiative, which are to expedite the development of advanced materials. Here show for the first time that it is possible to identify and produce organic solid materials with optimal nonlinear optical properties that make them excellent for THz generation.

**Acknowledgements**

We gratefully acknowledge the Simmons Research Fund for partial funding of this research. We also thank the National Science Foundation Division of Materials Research (DMR) for funding (grant # 2104317).

**References**

[1]     C. R. Groom, I. J. Bruno, M. P. Lightfoot, and S. C. Ward. The Cambridge Structural Database. Acta Crystallographica Section B **72**, 171 (2016).
 https://doi.org/10.1107/S2052520616003954

[2]     S. K. Kauwe, J. Graser, R. Murdock, and T. D. Sparks. Can machine learning find extraordinary materials? Computational Materials Science **174**, 109498 (2020).
 https://www.sciencedirect.com/science/article/pii/S0927025619307979

[3]     B. Meredig. Five High-Impact Research Areas in Machine Learning for Materials Science. Chemistry of Materials **31**, 9579 (2019).
 https://doi.org/10.1021/acs.chemmater.9b04078

[4]     R. J. Murdock, S. K. Kauwe, A. Y.-T. Wang, and T. D. Sparks. Is Domain Knowledge Necessary for Machine Learning Materials Properties? Integrating Materials and Manufacturing Innovation **9**, 221 (2020).
 https://doi.org/10.1007/s40192-020-00179-z

[5]     K. T. Schütt, H. E. Sauceda, P. J. Kindermans, A. Tkatchenko, and K. R. Müller. SchNet – A deep learning architecture for molecules and materials. The Journal of Chemical Physics **148**, 241722 (2018).
 https://doi.org/10.1063/1.5019779

[6]     A. Y.-T. Wang, S. K. Kauwe, R. J. Murdock, and T. D. Sparks. Compositionally restricted attention-based network for materials property predictions. npj Computational Materials **7**, 77 (2021).
 https://doi.org/10.1038/s41524-021-00545-1

[7]     T. Xie and J. C. Grossman. Crystal Graph Convolutional Neural Networks for an Accurate and Interpretable Prediction of Material Properties. Physical Review Letters **120**, 145301 (2018).
 https://link.aps.org/doi/10.1103/PhysRevLett.120.145301

[8]     S. Curtarolo, G. L. W. Hart, M. B. Nardelli, N. Mingo, S. Sanvito, and O. Levy. The high-throughput highway to computational materials design. Nature Materials **12**, 191 (2013).
 https://doi.org/10.1038/nmat3568

[9]     C. C. Fischer, K. J. Tibbetts, D. Morgan, and G. Ceder. Predicting crystal structure by merging data mining with quantum mechanics. Nature Materials **5**, 641 (2006).
 https://doi.org/10.1038/nmat1691

[10]    H. Koinuma and I. Takeuchi. Combinatorial solid-state chemistry of inorganic materials. Nature Materials **3**, 429 (2004).
 https://doi.org/10.1038/nmat1157

[11]    N. Marzari, A. Ferretti, and C. Wolverton. Electronic-structure methods for materials design. Nature Materials **20**, 736 (2021).
 https://doi.org/10.1038/s41563-021-01013-3

[12]    W. Sun, C. J. Bartel, E. Arca, S. R. Bauers, B. Matthews, B. Orvañanos, B.-R. Chen, M. F. Toney, L. T. Schelhas, W. Tumas, J. Tate, A. Zakutayev, S. Lany, A. M. Holder, and G. Ceder. A map of the inorganic ternary metal nitrides. Nature Materials **18**, 732 (2019).
 https://doi.org/10.1038/s41563-019-0396-2

[13]    R. M. Geilhufe, S. S. Borysov, A. Bouhon, and A. V. Balatsky. Data Mining for Three-Dimensional Organic Dirac Materials: Focus on Space Group 19. Scientific Reports **7**, 7298 (2017).
 https://doi.org/10.1038/s41598-017-07374-7

[14]     R. M. Geilhufe, S. S. Borysov, D. Kalpakchi, and A. V. Balatsky. Towards novel organic high-${T}_{c}$ superconductors: Data mining using density of states similarity search. Physical Review Materials **2**, 024802 (2018).
 https://link.aps.org/doi/10.1103/PhysRevMaterials.2.024802
[15]     R. M. Geilhufe, B. Olsthoorn, and A. V. Balatsky. Shifting computational boundaries for complex organic materials. Nature Physics **17**, 152 (2021).
 https://doi.org/10.1038/s41567-020-01135-6
[16]     W. L. Chan, J. Deibel, and D. M. Mittleman. Imaging with terahertz radiation. Reports on Progress in Physics **70**, 1325 (2007).
 http://stacks.iop.org/0034-4885/70/i=8/a=R02
[17]     J. B. Baxter and G. W. Guglietta. Terahertz Spectroscopy. Analytical Chemistry **83**, 4342 (2011).
 http://dx.doi.org/10.1021/ac200907z
[18]     A. A. Gowen, C. O'Sullivan, and C. P. O'Donnell. Terahertz time domain spectroscopy and imaging: Emerging techniques for food process monitoring and quality control. Trends in Food Science & Technology **25**, 40 (2012).
 http://www.sciencedirect.com/science/article/pii/S0924224411002937
[19]     W. Zouaghi, M. D. Thomson, K. Rabia, R. Hahn, V. Blank, and H. G. Roskos. Broadband terahertz spectroscopy: principles, fundamental research and potential for industrial applications. European Journal of Physics **34**, S179 (2013).
 http://stacks.iop.org/0143-0807/34/i=6/a=S179
[20]     I. F. Akyildiz, J. M. Jornet, and C. Han. Terahertz band: Next frontier for wireless communications. Physical Communication **12**, 16 (2014).
 http://www.sciencedirect.com/science/article/pii/S1874490714000238
[21]     F. D. J. Brunner, O. P. Kwon, S.-J. Kwon, M. Jazbinšek, A. Schneider, and P. Günter. A hydrogen-bonded organic nonlinear optical crystal for high-efficiency terahertz generation and detection. Optics Express **16**, 16496 (2008).
 http://www.opticsexpress.org/abstract.cfm?URI=oe-16-21-16496
[22]     T. Kampfrath, K. Tanaka, and K. A. Nelson. Resonant and nonresonant control over matter and light by intense terahertz transients. Nat Photon **7**, 680 (2013).
 http://dx.doi.org/10.1038/nphoton.2013.184
[23]     C. H. Matthias and F. József András. Intense ultrashort terahertz pulses: generation and applications. Journal of Physics D: Applied Physics **44**, 083001 (2011).
 http://stacks.iop.org/0022-3727/44/i=8/a=083001
[24]     C. Vicario, M. Jazbinsek, A. V. Ovchinnikov, O. V. Chefonov, S. I. Ashitkov, M. B. Agranat, and C. P. Hauri. High efficiency THz generation in DSTMS, DAST and OH1 pumped by Cr:forsterite laser. Optics Express **23**, 4573 (2015).
 http://www.opticsexpress.org/abstract.cfm?URI=oe-23-4-4573
[25]     G. A. Valdivia-Berroeta, L. K. Heki, E. W. Jackson, I. C. Tangen, C. B. Bahr, S. J. Smith, D. J. Michaelis, and J. A. Johnson. Terahertz generation and optical characteristics of P-BI. Optics Letters **44**, 4279 (2019).
 http://ol.osa.org/abstract.cfm?URI=ol-44-17-4279

[26]     G. A. Valdivia-Berroeta, L. K. Heki, E. A. McMurray, L. A. Foote, S. H. Nazari, L. Y. Serafin, S. J. Smith, D. J. Michaelis, and J. A. Johnson. Alkynyl Pyridinium Crystals for Terahertz Generation. Advanced Optical Materials **0**, 1800383 (2018).
 https://doi.org/10.1002/adom.201800383
[27]     G. A. Valdivia-Berroeta, E. W. Jackson, K. C. Kenney, A. X. Wayment, I. C. Tangen, C. B. Bahr, S. J. Smith, D. J. Michaelis, and J. A. Johnson. Designing Non-Centrosymmetric Molecular Crystals: Optimal Packing May Be Just One Carbon Away. Advanced Functional Materials **n/a**, 1904786 (2019).
 https://doi.org/10.1002/adfm.201904786
[28]     G. A. Valdivia-Berroeta, K. C. Kenney, E. W. Jackson, J. C. Bloxham, A. X. Wayment, D. J. Brock, S. J. Smith, J. A. Johnson, and D. J. Michaelis. 6MNEP: a molecular cation with large hyperpolarizability and promise for nonlinear optical applications. Journal of Materials Chemistry C  (2020).
 http://dx.doi.org/10.1039/D0TC01829E
[29]     I. C. Tangen, G. A. Valdivida-Berroeta, L. K. Heki, Z. B. Zaccardi, E. W. Jackson, C. B. Bahr, D. J. Michaelis, and J. A. Johnson. Comprehensive Characterization of Terahertz Generation with the Organic Crystal BNA. ArXiv e-prints, 2005.05545 (2020).

[30]     G. A. Valdivia-Berroeta, I. C. Tangen, C. B. Bahr, K. C. Kenney, E. W. Jackson, J. DeLagange, D. J. Michaelis, and J. A. Johnson. Crystal Growth, Tetrahertz Generation, and Optical Characterization of EHPSI-4NBS. The Journal of Physical Chemistry C **125**, 16097 (2021).
 https://doi.org/10.1021/acs.jpcc.1c01698
[31]     Z. B. Zaccardi, I. C. Tangen, G. A. Valdivia-Berroeta, C. B. Bahr, K. C. Kenney, C. Rader, M. J. Lutz, B. P. Hunter, D. J. Michaelis, and J. A. Johnson. Enabling High-Power, Broadband THz Generation with 800-nm Pump Wavelength. ArXiv e-prints, 2010.02380 (2020).

[32]     M. Shalaby, C. Vicario, K. Thirupugalmani, S. Brahadeeswaran, and C. P. Hauri. Intense THz source based on BNA organic crystal pumped at Ti:sapphire wavelength. Optics Letters **41**, 1777 (2016).
 http://ol.osa.org/abstract.cfm?URI=ol-41-8-1777
[33]     H. Zhao, Y. Tan, T. Wu, G. Steinfeld, Y. Zhang, C. Zhang, L. Zhang, and M. Shalaby. Efficient broadband terahertz generation from organic crystal BNA using near infrared pump. Applied Physics Letters **114**, 241101 (2019).
 https://doi.org/10.1063/1.5098855
[34]     C. B. Bahr, N. K. Green, L. K. Heki, E. McMurray, I. C. Tangen, G. A. Valdivia-Berroeta, E. W. Jackson, D. J. Michaelis, and J. A. Johnson. Heterogeneous layered structures for improved terahertz generation. Optics Letters **45**, 2054 (2020).
 http://ol.osa.org/abstract.cfm?URI=ol-45-7-2054
[35]     M. Jazbinsek, U. Puc, A. Abina, and A. Zidansek. Organic Crystals for THz Photonics. Applied Sciences **9** (2019).

[36]     J.-H. Jeong, B.-J. Kang, J.-S. Kim, M. Jazbinsek, S.-H. Lee, S.-C. Lee, I.-H. Baek, H. Yun, J. Kim, Y. S. Lee, J.-H. Lee, J.-H. Kim, F. Rotermund, and O. P. Kwon. High-power Broadband Organic THz Generator. Scientific Reports **3**, 3200 (2013).

http://dx.doi.org/10.1038/srep03200

[37]     O. P. Kwon, B. Ruiz, A. Choubey, L. Mutter, A. Schneider, M. Jazbinsek, V. Gramlich, and P. Günter. Organic Nonlinear Optical Crystals Based on Configurationally Locked Polyene for Melt Growth. Chemistry of Materials **18**, 4049 (2006).
http://dx.doi.org/10.1021/cm0610130

[38]     J.-A. Lee, W. T. Kim, M. Jazbinsek, D. Kim, S.-H. Lee, I. C. Yu, W. Yoon, H. Yun, F. Rotermund, and O. P. Kwon. X-Shaped Alignment of Chromophores: Potential Alternative for Efficient Organic Terahertz Generators. Advanced Optical Materials **8**, 1901921 (2020).
https://doi.org/10.1002/adom.201901921

[39]     S.-H. Lee, M. Jazbinsek, C. P. Hauri, and O. P. Kwon. Recent progress in acentric core structures for highly efficient nonlinear optical crystals and their supramolecular interactions and terahertz applications. CrystEngComm **18**, 7180 (2016).
http://dx.doi.org/10.1039/C6CE00707D

[40]     S.-H. Lee, B.-J. Kang, J.-S. Kim, B.-W. Yoo, J.-H. Jeong, K.-H. Lee, M. Jazbinsek, J. W. Kim, H. Yun, J. Kim, Y. S. Lee, F. Rotermund, and O. P. Kwon. New Acentric Core Structure for Organic Electrooptic Crystals Optimal for Efficient Optical-to-THz Conversion. Advanced Optical Materials **3**, 756 (2015).
http://dx.doi.org/10.1002/adom.201400502

[41]     L. Mutter, F. D. Brunner, Z. Yang, M. Jazbinšek, and P. Günter. Linear and nonlinear optical properties of the organic crystal DSTMS. Journal of the Optical Society of America B **24**, 2556 (2007).
http://josab.osa.org/abstract.cfm?URI=josab-24-9-2556

[42]     Z. Yang, L. Mutter, M. Stillhart, B. Ruiz, S. Aravazhi, M. Jazbinsek, A. Schneider, V. Gramlich, and P. Günter. Large-Size Bulk and Thin-Film Stilbazolium-Salt Single Crystals for Nonlinear Optics and THz Generation. Advanced Functional Materials **17**, 2018 (2007).
http://dx.doi.org/10.1002/adfm.200601117

[43]     Z. Yang, M. Wörle, L. Mutter, M. Jazbinsek, and P. Günter. Synthesis, Crystal Structure, and Second-Order Nonlinear Optical Properties of New Stilbazolium Salts. Crystal Growth & Design **7**, 83 (2007).
http://dx.doi.org/10.1021/cg060449p

[44]     J. Yin, L. Li, Z. Yang, M. Jazbinsek, X. Tao, P. Günter, and H. Yang. A new stilbazolium salt with perfectly aligned chromophores for second-order nonlinear optics: 4-N,N-Dimethylamino-4'-N'-methyl-stilbazolium 3-carboxy-4-hydroxybenzenesulfonate. Dyes and Pigments **94**, 120 (2012).
https://www.sciencedirect.com/science/article/pii/S0143720811003226

[45]     S. S. Borysov, R. M. Geilhufe, and A. V. Balatsky. Organic materials database: An open-access online database for data mining. PLOS ONE **12**, e0171501 (2017).
https://doi.org/10.1371/journal.pone.0171501